\documentclass[pre,twocolumn,showpacs,amsmath,amssymb]{revtex4}

\usepackage{mathrsfs}
\usepackage{epsfig} 
\usepackage{dcolumn}  
\usepackage{bm}       

\newcommand{\bsub}{\begin{subequations}}
\newcommand{\esub}{\end{subequations}}

\newcommand{\nablabf}{\boldsymbol{\nabla}}

\begin{document}

\title{Frequency response in surface-potential driven electro-hydrodynamics}

\author{L.~Ejsing, K.~Smistrup, C.~M. Pedersen, N.~A. Mortensen, and H.~Bruus}

\affiliation{MIC -- Department of Micro and Nanotechnology,\\
Technical University of Denmark, NanoDTU Bldg.~345 east, DK-2800
Kongens Lyngby, Denmark}

\date{February 9, 2006}
\begin{abstract}
Using a Fourier approach we offer a general solution to
calculations of slip velocity within the circuit description of
the electro-hydrodynamics in a binary electrolyte confined by a
plane surface with a modulated surface potential. We consider the
case with a spatially constant intrinsic surface capacitance where
the net flow rate is in general zero while harmonic rolls as well
as time-averaged vortex-like components may exist depending on the
spatial symmetry and extension of the surface potential. In
general the system displays a resonance behavior at a frequency
corresponding to the inverse RC time of the system. Different
surface potentials share the common feature that the resonance
frequency is inversely proportional to the characteristic length
scale of the surface potential. For the asymptotic frequency
dependence above resonance we find a $\omega^{-2}$ power law for
surface potentials with either an even or an odd symmetry. Below
resonance we also find a power law $\omega^\alpha$ with $\alpha$
being positive and dependent of the properties of the surface
potential. Comparing a $\tanh$ potential and a $\textrm{sech}$
potential we qualitatively find the same slip velocity, but for
the below-resonance frequency response the two potentials display
different power law asymptotics with $\alpha=1$ and $\alpha\sim
2$, respectively.
\end{abstract}

 \maketitle

\section{Introduction}

The ability to manipulate liquids on microfluidic chips is
essential to the functionality of micro total analysis
systems~\cite{Geschke:04a} and recently there has been quite some
interest in utilizing AC electrical surface potentials for
pumping, manipulating, and mixing electrolytes. For an overview of
electro-hydrodynamics and AC electro-osmosis we refer to
Refs.~\onlinecite{Ramos:98a,Ajdari:00a,Morgan:03a,Stone:04a,Bazant:04b,Squires:04a,Bazant:04a,Mortensen:05a}~and
references therein.

In this paper we consider the RC model in
Ref.~\onlinecite{Ajdari:00a} of an electrolyte confined by an
insulating surface with an applied external potential of the
general form
\begin{equation}\label{eq:Vext}
V_\textrm{ext}(y,t)=V_0 f(y) \exp(i\omega t).
\end{equation}
Here, $\omega$ is the driving frequency and $f(y)$ is a
dimensionless function representing the amplitude variations along
the surface. For simplicity we consider a binary electrolyte with
permeability $\epsilon$, viscosity $\eta$, and conductivity
$\sigma$. Assuming a low P\'eclet number we may neglect convection
so that the electrodynamics can formally be solved independently
of the hydrodynamics. On the other hand the hydrodynamics of
course still depends on the electrodynamics through the body force
which effectively may be taken into account through a finite slip
velocity, see e.g.
Refs.~\onlinecite{Gamayunov:86a,Murtsovkin:96a,Ajdari:00a,Green:00a,Squires:04a}
and references therein.

\section{Electrodynamics}
For the electrodynamics we follow Ajdari~\cite{Ajdari:00a} with
the surface and the Debye layer each represented by a capacitive
element while the bulk-liquid is represented by an ohmic element.
The model applies to the situation where spatial variations along
the surface are slow on the length scale of the Debye screening
length $\lambda_D$, the driving frequency is small compared to the
Debye frequency $\omega_D$, and the electrostatic energy is small
compared to the thermal energy $k_BT$, i.e. the Debye--H\"uckel
approximation is valid. As shown by many, the model may be
justified in detail starting from a non-equilibrium
electro-hydrodynamic continuum description. For more details we
refer to Refs.~\onlinecite{Gamayunov:86a,Murtsovkin:96a} as well
as~\onlinecite{Mortensen:05a,Bazant:04a} and references therein.

In the bulk of the liquid there is charge neutrality and the
electrical potential $\phi$ thus fulfills the Laplace equation
$\nablabf^2\phi=0$ subject to the relation between surface charge
$\sigma_D$ and the potential drop across the capacitive element
$\phi(0,y,t)\Big|-V_\textrm{ext}(y,t)=\sigma_D(y,t)/C_0$ as well
as a continuity equation
$\partial_t\sigma_D(y,t)=-J_x(0,y,t)=\sigma\partial_x
\phi(x=0,y,t)$ relating the change in surface charge to the
electrical Ohmic current $J$. Above,
$C_0=\big(C_s^{-1}+C_D^{-1}\big)^{-1}$ is a series capacitance
with $C_s$ being the intrinsic surface capacitance and
$C_D=\epsilon/\lambda_D$ being the Debye layer capacitance.

We formally solve this problem by introducing the inverse Fourier
transform ${\mathscr F}(Q)$ of $f(y)$, see Eq.~(\ref{eq:Vext}),
\begin{equation}
{\mathscr F}(Q)= \int_{-\infty}^\infty \frac{dy'}{\sqrt{2\pi}}\,
f(y') e^{-iQy'}.
 \end{equation}
From the solution of the spatially harmonic
problem~\cite{Ajdari:00a} we may write the general solutions for
the potential and the surface charge as
\begin{multline}\label{eq:phi}
\phi(x,y,t)= V_0\exp(i\omega t)\\\times \int_{-\infty}^\infty
\frac{dQ}{\sqrt{2\pi}}\, \frac{{\mathscr
F}(Q)\exp(iQy)\exp(-|Q|x)}
  {1  -i
   |Q|\Lambda_D }
\end{multline}
and
\begin{multline}\label{eq:sigmaD}
\sigma_D(y,t)= i\frac{\sigma V_0}{\omega}\exp(i\omega
t)\\\times\int_{-\infty}^\infty \frac{dQ}{\sqrt{2\pi}}\, \frac{
{\mathscr F}(Q)|Q|\exp(iQy)}{1  -i |Q|\Lambda_D}.
\end{multline}
Above, we have introduced the frequency dependent effective length
scale $\Lambda_D$ given by
\begin{equation}
\Lambda_D=(1+\delta)\frac{\omega_D}{\omega}\:\lambda_D,
\end{equation}
where $\delta= \frac{C_D}{C_s}$ is the capacitance ratio and
$\omega_D=\frac{D}{\lambda_D^2}$ is the Debye frequency.

\section{The slip velocity}

The slip velocity is given by~\cite{Ajdari:00a}
\begin{equation}
v_\textrm{slip}(y)=\frac{\lambda_D}{\eta}\sigma_D(y)\Big[-
\partial_y\phi(x,y)\Big]_{x=0},
\end{equation}
where it is implicit that real values of $\sigma_D$ and $\phi$
must be taken. From Eqs.~(\ref{eq:phi}) and (\ref{eq:sigmaD}) we
now formally get
 \begin{align}
 & v_\textrm{slip}(y,t)= -\frac{\lambda_D\epsilon
 V_0^2}{\eta}\frac{\omega_D}{\omega}
 \nonumber\\ & \qquad \times
 \textrm{Re}\Bigg\{ie^{i\omega t}\int_{-\infty}^\infty \frac{dQ}
 {\sqrt{2\pi}}\, \frac{ {\mathscr F}(Q)|Q|\exp(iQy)}{1  -i
 |Q|\Lambda_D}\Bigg\}
 \nonumber \\ & \qquad \times
 \textrm{Re}\Bigg\{ie^{i\omega t} \int_{-\infty}^\infty \frac{dQ'}
 {\sqrt{2\pi}}\,\frac{{\mathscr F}(Q')Q' \exp(iQ'y)}
 {1  -i |Q'|\Lambda_D }\Bigg\}, \label{eq:vs}
 \end{align}
and for the time-averaged slip-velocity
$\big<v_\textrm{slip}\big>_t$ we use that $\big< \textrm{Re}\{f
e^{i\omega t}\}\textrm{Re}\{g e^{i\omega t}\} \big>_t =
\frac{1}{2}\textrm{Re}\{f^*g\}$ as well as ${\mathscr
F}^*(Q)={\mathscr F}(-Q)$ so that
 \begin{multline}\label{eq:vs_t}
 \big<v_\textrm{slip}(y,t)\big>_t= \frac{\lambda_D\epsilon
 V_0^2}{2\eta}\frac{\omega_D}{\omega}
 \textrm{Re}\Bigg\{\int_{-\infty}^\infty \frac{dQ}{\sqrt{2\pi}}\,
 \int_{-\infty}^\infty \frac{dQ'}{\sqrt{2\pi}}\,\\ \times
 \frac{ {\mathscr F}(Q)|Q|\exp(iQy)}{1  -i |Q|\Lambda_D}
 \:\frac{{\mathscr
  F}(Q')Q' \exp(iQ'y)}
  {1  +i
   |Q'|\Lambda_D }\Bigg\}.
\end{multline}
Eqs.~(\ref{eq:vs}) and (\ref{eq:vs_t}) are our general results for
the surface-potential induced hydrodynamics. Velocity and pressure
fields may be obtained by solving the Navier--Stokes equation
without body-forces subject to a slip-velocity condition.

Before turning our attention to specific potentials we may already
extract some general properties of the model. From
Eq.~(\ref{eq:vs_t}) it is clear that
$\big<v_\textrm{slip}(y,t)\big>_t$ is a spatially dependent
function of finite magnitude. This in turn also means that the
external potential can induce a finite time-averaged velocity
field in the fluid. However, it is also clear that the induced net
flow rate will be zero since $\big<v_\textrm{slip}(y,t)\big>_t$
has no spatial dc-component, as may be seen by substituting the
general expression for ${\mathscr F}(Q)$. Spatial asymmetry in the
applied potential is thus not a sufficient condition to achieve
pumping and as also indicated in Ref.~\onlinecite{Ajdari:00a},
asymmetry in the surface capacitance is required. Formal studies
of the model for $\partial_y C_s\neq 0$ will be an interesting
subject for future studies, but here we will follow the lines were
$C_s$ is constant along the surface. The case of $f(y)=\cos(qy)$
was first studied in Ref.~\onlinecite{Ajdari:00a}. The Fourier
transform is then a simple sum of two Dirac delta functions,
${\mathscr F}(Q)\propto \delta(Q-q)+\delta(Q+q)$, and substituting
into Eqs.~(\ref{eq:vs}) or (\ref{eq:vs_t}) we easily arrive at
results fully consistent with those reported previously.

\begin{figure*}[t!]
\begin{center}
\epsfig{file=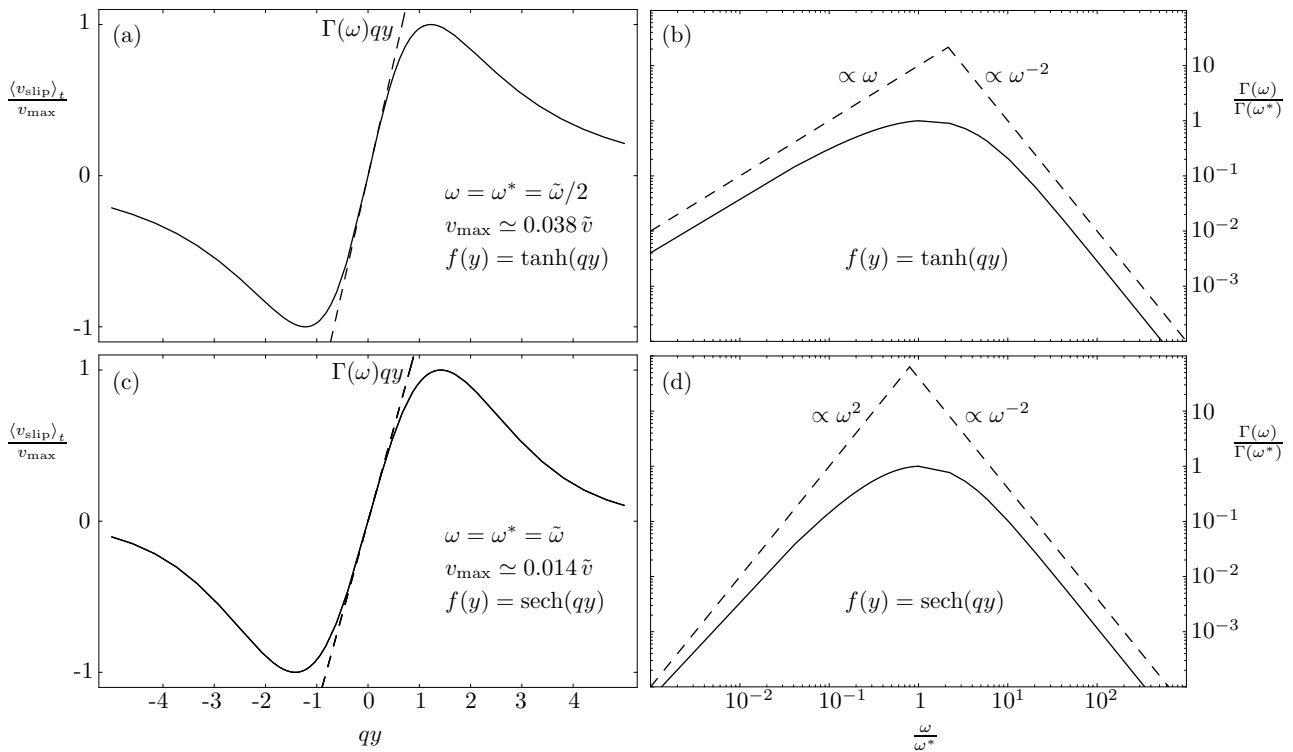, width=0.95\textwidth,clip}
\end{center}
\caption{Dynamics for $\tanh(qy)$ and ${\rm sech}(qy)$ surface
potentials, panels (a)-(b) and (c)-(d), respectively. Panels (a)
and (c) show the time-averaged slip velocities at resonance while
panels (b) and (d) show the corresponding frequency response.
In~(a) and~(c) the solid lines show numerical evaluations of
Eqs.~(\ref{eq:vs_t_tanh}) and~(\ref{eq:vs_t_sech}), respectively,
while dashed lines show the slope $\partial_y \big< v_{\rm
slip}\big>_t$ at $y=0$ calculated from Eqs.~(\ref{eq:Gamma_tanh})
and~(\ref{eq:Gamma_sech}), respectively. In (b) and (d) the solid
lines show numerical evaluations of Eqs.~(\ref{eq:Gamma_tanh})
and~(\ref{eq:Gamma_sech}), respectively, while dashed lines are
included to indicate asymptotics.} \label{fig1}
\end{figure*}

\section{Frequency dynamics}

In order to draw some general conclusions about the frequency
dynamics of the hydrodynamics, apart from the frequency doubling
\cite{Ajdari:00a}, we imagine that the surface potential has a
characteristic length scale $1/q$. Since we have made an RC
description of the electro-hydrodynamics it follows quite
naturally that the system will have a resonance behavior at a
frequency comparable to the inverse RC time of the system. Above
resonance the system has an asymptotic power-law frequency
response as we will illustrate below for a surface potential of
even symmetry. In that case ${\mathscr F}(Q)$ is a real and even
function of $Q$, so that Eq.~(\ref{eq:vs_t}) with $Q=sq$ becomes
 \begin{align}
 \big<v_\textrm{slip}(y,t)\big>_t =&\; 2\tilde{v} q^2
 \textrm{Re}\Bigg\{\int_{0}^\infty \frac{ds}{\sqrt{2\pi}}\, \frac{
 {\mathscr F}(s q)s \cos(s qy)}{\frac{\omega}{\tilde\omega} -i s }
 \nonumber \\ &
 \times\int_{0}^\infty \frac{ds'}{\sqrt{2\pi}}\,
 \frac{i\frac{\omega}{\tilde\omega}{\mathscr F}(s'q)s' \sin(s'qy)}
  {\frac{\omega}{\tilde\omega}  +i s'}\Bigg\},
 \end{align}
where we have introduced a characteristic resonance
frequency~$\tilde\omega$ as well as a characteristic flow velocity
$\tilde{v}$,
 \begin{equation}
 \tilde\omega\equiv q\lambda_D(1+\delta)\omega_D
 = q\Lambda_D\omega,\quad
\tilde{v}= \frac{q\epsilon V_0^2}{\eta(1+\delta)}.
\end{equation}
Since
\begin{align}\nonumber
\textrm{Re}\Bigg\{ \frac{1}{\Omega  -i s }
   \frac{i\Omega}
  {\Omega  +i
   s'}\Bigg\}&=\frac{(s'-s)\Omega^2}{[\Omega^2+s^2][\Omega^2+(s')^2]}\\&=(s-s')
   \Omega^{-2}+{\cal
   O}\left(\Omega^{-4}\right)
\end{align}
we expect with $\Omega=\omega/\tilde\omega$ that
\begin{equation}\label{eq:aboveresonance}
\big<v_\textrm{slip}\big>_t \propto \omega^{-2},\quad
\omega\gg\tilde\omega.
\end{equation}
Likewise, for surface potentials of odd symmetry we arrive at the
same conclusion. Making a similar expansion for the
below-resonance asymptote one might speculate that
$\big<v_\textrm{slip}\big>_t \propto \omega^\alpha$ with a power
$\alpha=2$. However, with such an expansion the integrals are in
general no longer convergent which in turn indicates that the
power $\alpha$ is below 2 depending on the actual surface
potential under consideration. In the following we will illustrate
this by the aid of two explicit examples.

\subsection{Two-electrode configuration}

Recently, a circulating flow was observed experimentally using two
adjacent electrodes and an AC field \cite{Green:00a} and in Ref.
\cite{Gonzalez:00a} the problem was studied theoretically by
assuming that the two electrodes are infinitely close
corresponding to a step-like model potential. However, such a
potential is somewhat unphysical since the electric field diverges
and instead one should take the electrode gap into account
~\cite{Ramos:03a}. In the following we model such an electrode
configuration with a smooth external potential of the form
$f(y)=\tanh(qy)$ which incorporates the built-in electrode gap in
terms of the length scale $1/q$. Fourier transforming and
substituting into Eq.~(\ref{eq:vs_t}) we get
\begin{multline}\label{eq:vs_t_tanh}
\big<v_\textrm{slip}(y,t)\big>_t=\tilde{v}
\textrm{Re}\Bigg\{\int_{0}^\infty ds\, \frac{
\textrm{csch}\big(\frac{\pi }{2}s\big)s\sin(s
qy)}{i\frac{\omega}{\omega^*} + 2s}
  \\\times\int_{0}^\infty ds'\, \frac{i\frac{\omega}{\omega^*}\:\textrm{csch}\big(\frac{\pi
}{2}s'\big)s' \cos(s' qy)}
  {i\frac{\omega}{\omega^*}  -2s' }\Bigg\},
\end{multline}
where the resonance frequency is given by
$\omega^*=\tilde{\omega}/2$. In order to study the maximal value
of $\big<v_\textrm{slip}(y,t)\big>_t$ as a function of frequency
we note that when $\big<v_\textrm{slip}(y,t)\big>_t$ is maximal so
is its slope at $y=0$. If we define
$\Gamma(\omega)=(q\tilde{v})^{-1}\partial_y\big<v_\textrm{slip}(y,t)\big>_t\big|_{y=0}$
we get
\begin{equation}\label{eq:Gamma_tanh}
\Gamma=-2\Omega^2(J_2^2-J_1J_3),\quad J_n = \int_0^\infty
ds\,\frac{ \textrm{csch}(\frac{\pi}{2}s)s^n}{\Omega^2 +
   4s^2}
\end{equation}
where we have introduced $\Omega=\omega/\omega^*$. Panel (a) of
Fig.~\ref{fig1} shows the time-averaged slip velocity at resonance
evaluated numerically from Eq.~(\ref{eq:vs_t_tanh}) and its slope
at $y=0$, based on a numerical evaluation of
Eq.~(\ref{eq:Gamma_tanh}), is also included. Panel (b) shows
$\Gamma(\omega)$ as a function of frequency. The plot confirms
Eq.~(\ref{eq:aboveresonance}) and suggests a linear dependence
$\big<v_\textrm{slip}\big>_t\propto\omega$ for the below-resonance
frequency response. This may also confirmed by an asymptotic
analysis by first noting that $J_1=1/(2\Omega)+{\cal
O}(\Omega^2)$. For
   $J_2$ we note that $\partial J_2/\partial\Omega =-2\Omega \int_0^\infty ds\, \textrm{csch}(\frac{\pi}{2}s)s^2/[\Omega^2 +
   4s^2]^2\simeq -4\Omega/\pi \int_0^\infty ds\, s/[\Omega^2 +
   4s^2]^2=-1/2\pi\Omega$ so that $J_2\simeq {\rm
   const.}-\ln(\Omega)/2\pi$. For $J_3$ we similarly note that $\partial J_3/\partial\Omega\simeq -1/8$ so that $J_3={\rm
   const.}-\Omega/8$. When collecting terms we thus get a
   $\omega$-dependence to lowest order.

\subsection{Single-electrode configuration}
As a second example we consider single-electrode driven
electro-hydrodynamics and as a model potential we use
$f(y)=\textrm{sech}(qy)$ to mimic the potential of a single
electrode with a characteristic width $1/q$. Fourier transforming
and substituting into Eq.~(\ref{eq:vs_t}) we get
\begin{multline}\label{eq:vs_t_sech}
\big<v_\textrm{slip}(y,t)\big>_t= -\frac{1}{2}\tilde{v}
\textrm{Re}\Bigg\{\int_0^\infty ds\,
\frac{i\frac{\omega}{\omega^*}\:
\textrm{sech}(\frac{\pi}{2}s)s\cos(s\:
qy)}{i\frac{\omega}{\omega^*} + s }
  \\\times\int_{0}^\infty ds'\, \frac{ \textrm{sech}(\frac{\pi}{2}s')s'
\sin(s'qy)}
  {i\frac{\omega}{\omega^*} -
   s' }\Bigg\},
\end{multline}
where in this case $\omega^*=\tilde\omega$ is the resonance
frequency. In analogy with the previous example we get
\begin{equation}\label{eq:Gamma_sech}
\Gamma=-\frac{1}{2}\Omega^2(I_2^2-I_1I_3),\quad I_n =
\int_0^\infty ds\,
\frac{\textrm{sech}(\frac{\pi}{2}s)s^n}{\Omega^2 +
   s^2}
\end{equation}
Panel (c) in Fig.~\ref{fig1} shows the time-averaged slip velocity
at resonance evaluated numerically from Eq.~(\ref{eq:vs_t_sech})
and the slope at $y=0$ based on a numerical evaluation of
Eq.~(\ref{eq:Gamma_sech}) is also included. Panel (d) shows the
frequency response confirming Eq.~(\ref{eq:aboveresonance}) for
the above-resonance response while for the below-resonance
response $\big<v_\textrm{slip}\big>_t\propto\omega^\alpha$ with a
power $\alpha\sim 2$. For the asymptotic analysis we note that
$I_2= 1+{\cal O}(\Omega^2)$ and $I_3= 8 C/\pi^2+{\cal
O}(\Omega^2)$ where
   $C\simeq 0.915966$ is Catalan's constant. For $I_1$ it follows that $\partial I_1/\partial\Omega =-2\Omega \int_0^\infty ds\, \textrm{sech}(\frac{\pi}{2}s)s/[\Omega^2 +
   s^2]^2 \simeq -2\Omega \int_0^\infty ds\, s/[\Omega^2 +
   s^2]^2= - \Omega^{-1}$
   so that $I_1={\rm const.}-\ln(\Omega)+{\cal O}(\Omega^2)$.
   There will thus be a logarithmic correction $\omega^2 \ln(\omega)$ to the $\omega^2$
dependence.

\section{Discussion and conclusion}

In this work we have used a Fourier transform approach to make a
general analysis of the frequency dynamics of AC-induced
electro-hydrodynamics within the RC model of
Ref.~\onlinecite{Ajdari:00a}. In the case with a spatially
constant intrinsic surface capacitance we generally prove that the
net flow rate is zero. A finite net flow rate will require a
spatially inhomogeneous capacitance so that the charge and
electrical field dynamics are mutually out of phase.

For the frequency dynamics we observe a resonance behavior at a
frequency corresponding to the inverse RC time of the system and
different surface potentials share the common feature that the
resonance frequency is inversely proportional to the
characteristic length scale of the surface potential. Above
resonance we in general find a $\omega^{-2}$ power law for surface
potentials with either even or odd symmetry. Below resonance we
likewise find a power-law dependence $\omega^\alpha$ with $\alpha$
being positive and dependent of the properties of the surface
potential. As two examples we have compared $\tanh$ (odd symmetry)
and $\textrm{sech}$ (even symmetry) surface potential.
Unexpectedly, we qualitatively find the same slip velocity, while
the frequency dynamics is very different for the below-resonance
frequency response where power law asymptotics with $\alpha=1$ and
$\alpha\sim 2$, respectively, are found.

Our study illustrates how the frequency dynamics carries strong
fingerprints of the driving surface potential and we believe that
detailed analysis of the frequency dynamics would be a way to
extract more information from experimental data. Recent related
work~\cite{Levitan:05a} also suggests the possibility for a strong
frequency dependence in ac electroosmosis also in the flow
topology.

\section*{Acknowledgement}

We thank L.~H.~Olesen for stimulating discussions. K.~S. is
supported by C:O:N:T and Teknologisk Institut. N.~A.~M. is
supported by The Danish Technical Research Council (Grant
No.~26-03-0073).


\end{document}